# Practice of Alibaba Cloud on Elastic Resource Provisioning for Large-scale Microservices Cluster

| | |
|---|---|
| Journal: | *Software: Practice and Experience* |
| Manuscript ID | SPE-23-0220.R1 |
| Wiley - Manuscript type: | Research Article |
| Date Submitted by the Author: | 28-Jul-2023 |
| Complete List of Authors: | Xu, Minxian; Chinese Academy of Sciences Shenzhen Institutes of Advanced Technology<br>Yang, Lei; Chinese Academy of Sciences Shenzhen Institutes of Advanced Technology<br>Wang, Yang; Chinese Academy of Sciences Shenzhen Institutes of Advanced Technology<br>Gao, Chengxi; Chinese Academy of Sciences Shenzhen Institutes of Advanced Technology<br>Wen, Linfeng; Chinese Academy of Sciences Shenzhen Institutes of Advanced Technology<br>Xu, Guoyao; Alibaba Group<br>Zhang, Liping; Alibaba Group<br>Ye, Kejiang; Chinese Academy of Sciences Shenzhen Institutes of Advanced Technology<br>Xu, Chengzhong; University of Macau |
| Subject Area: | Distributed Computing Paradigms: Service, Cloud, Fog, Edge |
| Keywords: | Cloud-native, Microservice, Alibaba, Resource Provisioning, Latency |
| | |

# Practice of Alibaba Cloud on Elastic Resource Provisioning for Large-scale Microservices Cluster

Minxian Xu[1] | Lei Yang[1] | Yang Wang[1] | Chengxi Gao[1] | Linfeng Wen[1] | Guoyao Xu[2] | Liping Zhang[2] | Kejiang Ye[1] | Chengzhong Xu[3]

[1]Shenzhen Institutes of Advanced Technology, Chinese Academy of Sciences, Shenzhen, China
[2]Alibaba Group, Hangzhou, China
[3]State Key Lab of IOTSC, Department of Computer Science, University of Macau, Macau SAR, China

**Correspondence**
Minxian Xu (mx.xu@siat.ac.cn) and Lei Yang (lei.yang2@siat.ac.cn) are co-first authors ; Kejiang Ye (kj.ye@siat.ac.cn) and Chengzhong Xu (czxu@um.edu.mo) are corresponding authors.

**Summary**

Cloud-native architecture is becoming increasingly crucial for today's cloud computing environments due to the need for speed and flexibility in developing applications. It utilizes microservice technology to break down traditional monolithic applications into light-weight and self-contained microservice components. However, as microservices grow in scale and have dynamic inter-dependencies, they also pose new challenges in resource provisioning that cannot be fully addressed by traditional resource scheduling approaches. The various microservices with different resource needs and latency requirements can create complex calling chains, making it difficult to provide fine-grained and accurate resource allocation to each component while maintaining the overall quality of service in the chain. Alibaba Cloud has fully embraced cloud-native and microservice technologies to drive its key business and scenarios, including Double 11 Shopping Festival. In this work, we aim to address the research problem on how to efficiently provision resources for the growing scale of microservice platform and ensure the performance of latency-critical microservices. To address the problem, we present in-depth analyses of Alibaba's microservice cluster and propose optimized resource provisioning algorithms to enhance resource utilization while ensuring the latency requirement. First, we analyze the distinct features of microservices in Alibaba's cluster compared to traditional applications. Then we present Alibaba's resource capacity provisioning workflow and framework to address challenges in resource provisioning for large-scale and latency-critical microservice clusters. Finally, we propose enhanced resource provisioning algorithms over Alibaba's current practice by making both proactive and reactive scheduling decisions based on different workloads patterns, which can improve resource usage by 10-15% in Alibaba's clusters, while maintaining the necessary latency for microservices.

# 1 INTRODUCTION

Cloud-native architecture and related technologies aim to build and run scalable applications in modern and dynamic cloud environments by taking full advantage of the cloud service model[1]. Key concerns of technologies are speed and agility . Business systems require immediate transformation to accelerate business velocity and expand market growth. Meanwhile, the complexity of a business system tends to increase dramatically with the users' demand on rapid innovative features and rapid responsiveness.

Cloud-native architecture supports the service providers to embrace rapid change, resilience, and large scale, thus significant changes can be made frequently with minimal efforts. Prominent companies such as Netflix, Uber, and WeChat, have adopted cloud-native technologies, where hundreds of services in production can be deployed or updated 100 to 1000 times per day[2]. Key pillar that supports the speed and agility of cloud-native is microservice architecture[3]. The microservice has shifted traditional monolithic applications into loosely-coupled, light-weight, and self-contained microservice components. Each microservice unit can be deployed and operated independently for different business objectives and functionalities. Moreover, the microservices can interact with each other and collaborate as a whole application through light-weight communication. Through leveraging a set of independent and light-weight microservice components, microservice architecture has significantly improved the efficiency of application development and deployment[4 5].

The growing scale of microservice platforms and dynamic inter-dependencies between microservices raise new challenges in resource provisioning of cloud infrastructure[6]. The challenges include 1) multiple microservice units can form calling chains with varied lengths to fulfill complicated functionalities (e.g. complex business transactions). The dependencies between the microservice units make it challenging to investigate the relationship between the provisioned resources and quality of service (QoS), such as identifying the root microservice unit that causes QoS degradation. 2) Microservice can be much more latency-sensitive compared with a monolithic application, and the microservice units in the same chain can have varied latency and resource requirements, which make the independent resource provisioning to the units difficult. For instance, provisioning more resources to an under-utilized unit in the chain cannot mitigate the service delay of the microservice chain. 3) Co-locating different microservices can reduce costs while also incurring a dramatic change in resource usage due to workload variance, which can lead to high latency if resources are not provisioned timely.

Apart from the above challenges in resource provisioning, the industry also raises challenges due to the management of large-scale microservice clusters. For instance, Alibaba has migrated its services to cloud-native clusters and comprehensively applied microservice-based technologies[7]. The scale of Alibaba's cloud-native clusters has reached 10 million cores in 2021, and Alibaba System Infrastructure (**ASI**) has successfully supported China Double 11 Shopping Festival with 0.58 million transactions per second during the peak time. At this scale of a cluster in a production environment, the ordinary resource provisioning approaches validated in a research environment can be ineffective due to the power of scale. For instance, it is infeasible to explore all the possible solution space with pressure tests in a production environment concerning the huge amount of configurable parameters.

To address the challenges in resource provisioning of production environment, in this paper, we present resource management approaches of Alibaba's current practice for large-scale microservice cluster. The approaches have successfully supported the key businesses of Alibaba with high elasticity and plasticity. We also propose enhanced resource provisioning algorithms to optimize resource usage over Alibaba's practice while ensuring latency requirement.

The main **contributions** of this paper are as below:

- We present the architecture of Alibaba's microservice cluster designed to handle large-scale microservice management, along with comprehensive statistical analysis of the microservices in its production environment (Section 3).
- We provide key design of Alibaba's general resource provisioning framework in current practice (Section 4).
- We propose enhanced resource allocation methods that build upon Alibaba's current practices to efficiently and elastically support services through various means such as workload estimation, capability modeling, and resource allocation policies (Section 5).
- We evaluate our resource provisioning approach and show that it can increase resource utilization by 10-15% while maintaining QoS (Section 6).

The rest of this paper is organized as follows: the related work is discussed in Section 2. Section 3 presents the background information of Alibaba's cluster to support its scenario and the analyses of the features of its microservices along with the identified key challenges in resource provisioning. The current resource capacity provisioning practice of Alibaba and our enhanced solutions are proposed in Section 4 and Section 5 respectively. Section 6 demonstrates the conducted experiments in the Alibaba cluster and the conclusions and future work are given in Section 7.

## 2 | RELATED WORK

Resource capacity provisioning for cloud applications is a popular topic, and we categorize prior work into two buckets for traditional cloud applications and modern microservice applications.

**Resource Capacity Provisioning for Traditional Cloud Applications.** Prior research has proposed a set of resource capacity provisioning approaches for traditional application deployed on virtual machines[8 9]. Kiani et al.[10] proposed a hierarchical capacity provisioning scheme that applies a two-tier network architecture with shallow and deep cloudlets. Two models for network delay scenario with bufferless and finite-size buffer shallow cloudlets are modeled. Stochastic ordering is also applied to solve the optimization problem formulated for the models. Ma et al.[11] proposed a cloud-assisted mobile edge computing framework to optimize the computation capacity of edge hosts, and investigated the resource capacity provisioning problem with dynamic requests and proves it to be piecewise convex, which can determine the optimal computation capacity of edge hosts. Aslanpour et al.[12] presented an automatic resource provisioning approach considering resource usage, SLA, and user behavior. The proposed approach applies radial basis function neural networks to ensure dynamic resource provisioning. Xu et al.[13] proposed a resource provisioning approach to improve system fault tolerance for data-intensive meteorological workflows. The approach exploits virtual layer network topology and genetic algorithm to optimize makespan and balance loads. However, these approaches are designed for traditional monolithic applications or virtual machines in cloud computing environment, which are not suitable for the new features of microservices such as fine-grained and time-sensitive. In addition, the structure of microservices are not considered in these approaches, thus utilizing these approaches without modification can undermine the performance of microservices.

**Resource Capacity Provisioning for Microservices.** More attention has been paid to resource capacity provisioning for microservice applications under cloud-native environment[14,6,15]. Yu et al.[16] presented an online approach to dynamically identify microservices needed to be scaled and provision resources for microservices to assure the quality of services. The online learning approach and heuristic method have been utilized to achieve an optimal amount of scaled resources and they can achieve a faster convergence rate than baselines. Abdulla et al.[15] introduced a burst-aware auto-scaling approach to detect burst in dynamic workloads by workloads prediction and resource usage estimation. The scaling decisions for microservice can significantly reduce service level objective violations based on evaluations on synthetic and realistic bursty workloads for microservices. Zhang et al.[17] proposed data-driven cluster management for online and QoS-aware microservices. A set of machine learning (ML) models are applied to determine the appropriate resources to preserve end-to-end tail latency target. Sinan has been evaluated in experimental microservice cluster to improve utilization while ensuring QoS. Hou et al.[18] presented a power management framework for microservices focusing on decreasing power consumption of microservices cluster while reducing configuration latency by coordinating both macro and micro resource provisioning. The coordinated framework provides necessary abstraction and optimization based on resource provisioning to achieve better trade-offs between performance and power can be investigated. Kwan et al.[19] proposed a hybrid scaling approach combining vertical scaling and horizontal scaling for microservices. The proposed approach can leverage both the high availability of horizontal scaling and fine-grained control on resources of vertical scaling. Xu et al.[20] proposed deep learning based approach for cloud workloads prediction and reinforcement learning based approach to make auto-scaling decisions for microservices. These approaches have advanced the research area of resource capacity provisioning for microservices, however, they are mainly based on small-scale experience and cannot represent the practice in production environments.

**Large-scale Microservice Cluster Management.** Some industry articles have demonstrated their practice in production environment[21]. Newell et al.[22] described Facebook's region-scale resource scheme to dynamically assign servers to containers based on reservation to address correlated hardware failures, which has been utilized to the scale of millions of servers. Zhou et al.[23] managed large-scale WeChat microservices with overload control. The overload control scheme monitors load status in real-time and determines load shedding to match the upstream and downstream services. The scheme can also adjust resource provisioning to match the upstream and downstream services. Luo et al.[24] provided the analyses of the deployed microservice in Alibaba's production environment with focus of graph-based dependency. However, these work focuses on addressing microservice failures recovery, load balancing and dependency characterization rather than elastic capacity provisioning. In this paper, we present comprehensive statistical data of microservice behavior and general resource provisioning framework design of Alibaba's production environment. Moreover, we also propose optimized policies for Alibaba's platform to optimize resource usage while assuring QoS.

# 3 | BACKGROUND AND CHALLENGES

In this section, we present a comprehensive overview of Alibaba's microservice clusters, including its architecture, scale, and key features. The clusters were able to successfully handle the high demand of China's Double 11 Shopping Festival, reaching a peak of 0.58 million transactions per second.

## 3.1 | Architecture of Alibaba Microservice Clusters

Alibaba aims to provide uniform infrastructure for various microservice applications with heterogeneous hardware, therefore, Alibaba has developed the infrastructure, named ASI, based on Kubernetes[1] as shown in Fig. 1. A representative project derived from the infrastructure is PouchContainer to provide applications with a light-weight runtime environment with strong isolation and minimal overhead[25]. To satisfy the development of Alibaba's business and daily maintenance, a set of add-on extensions have been integrated to improve the capability of application deployment and management. The containerized pods are deployed on X-Dragon bare metals, which can support virtualization of computing, network, and storage resources. The X-Dragon physical machines exploit field-programmable gate array (FPGA) acceleration technology to enhance the performance of data communication and storage by reducing virtualization costs.

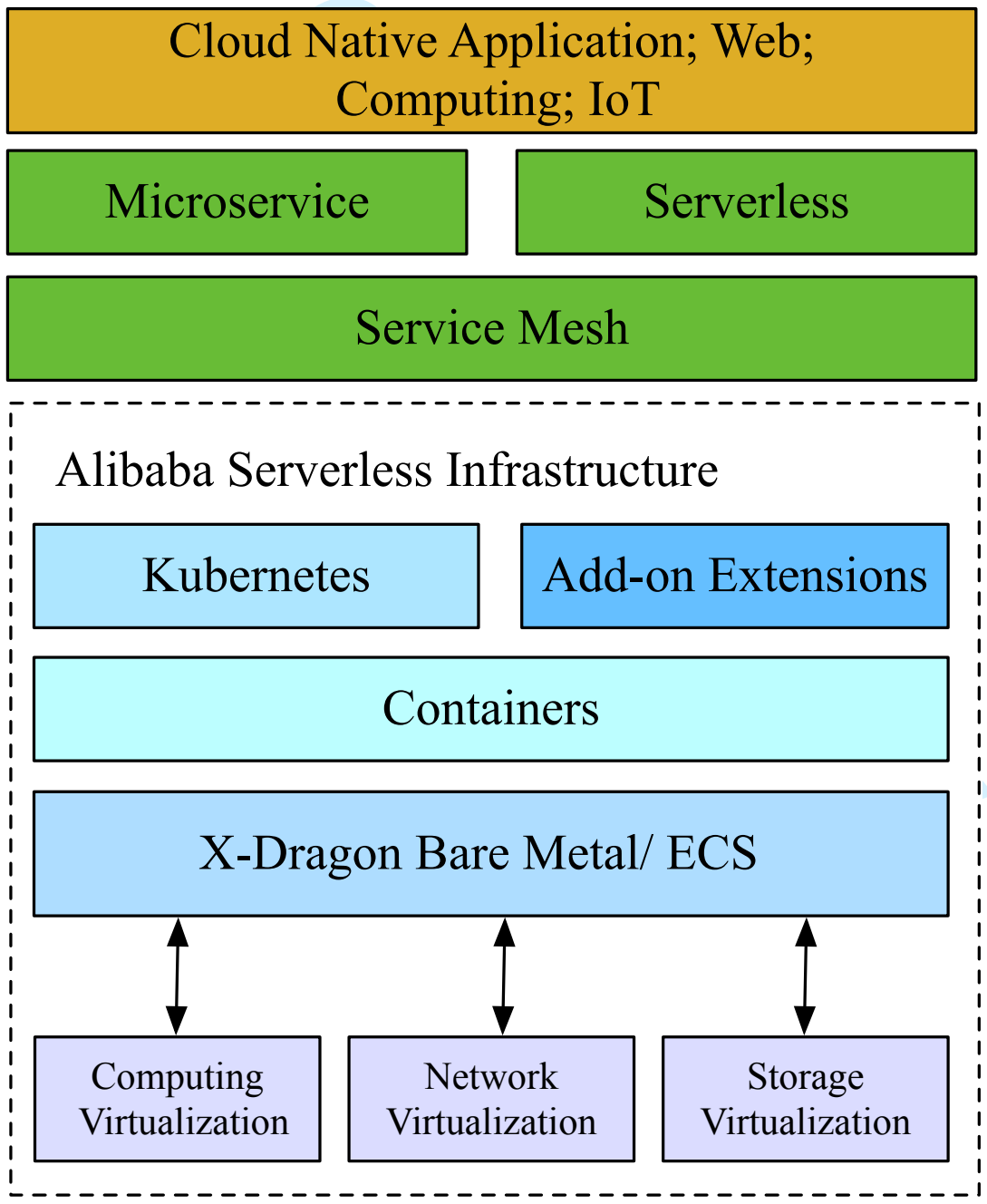

**Figure 1** Microservice architecture of Alibaba.

Above ASI, cloud-native applications including microservice, service mesh, and serverless can all be supported. Taking full advantage of the cloud platform, ASI can support tasks with different service level objectives (SLOs), including long lifetime pod replicates, batch tasks, computation-intensive tasks, and I/O intensive tasks. Alibaba has mainly categorized its services into three types with different priorities in resource capacity provisioning: 1) *Product Services* (e.g. critical and long lifetime services for business transactions) allocated with a specific budget quota and the resource capacity provisioner must assure the service availability with the highest priority. These services require strict computation resources, high reliability, low delay, and uninterrupted assurance. 2) *Batch Services* (e.g. document processing services) that are insensitive to the response time and has non-strict quota, while high availability should still be ensured. 3) *Best Effort Services* generally use the resources left by product and batch services and can be interrupted.

[1] https://kubernetes.io/

## 3.2 | Scale and Capability of Alibaba Microservice Clusters

The ASI is currently comprised of 50 large-scale clusters, each capable of hosting over 10,000 nodes and 10 million cores. The ASI supports around 10,000 applications and 100,000 business pods, serving all key business scenarios, including online sales, online video, mapping, food delivery, and virtual business meetings. The usage patterns of Alibaba's applications fluctuate based on user behavior. For example, food delivery applications experience high demand during mealtime and online shopping applications may see double the workload[2] during peak periods compared to off-peak periods[26]. The difference in queries per second (QPS) between peak and off-peak times for backend applications can be as high as 10 times. During major events, such as the Double 11 Shopping Festival, the number of transactions can reach ten times the ordinary workload. In 2019, the number of sale transactions reached 544,000 per second, and in 2020, it grew to 583,000 per second, a 1400% increase compared to ten years prior.

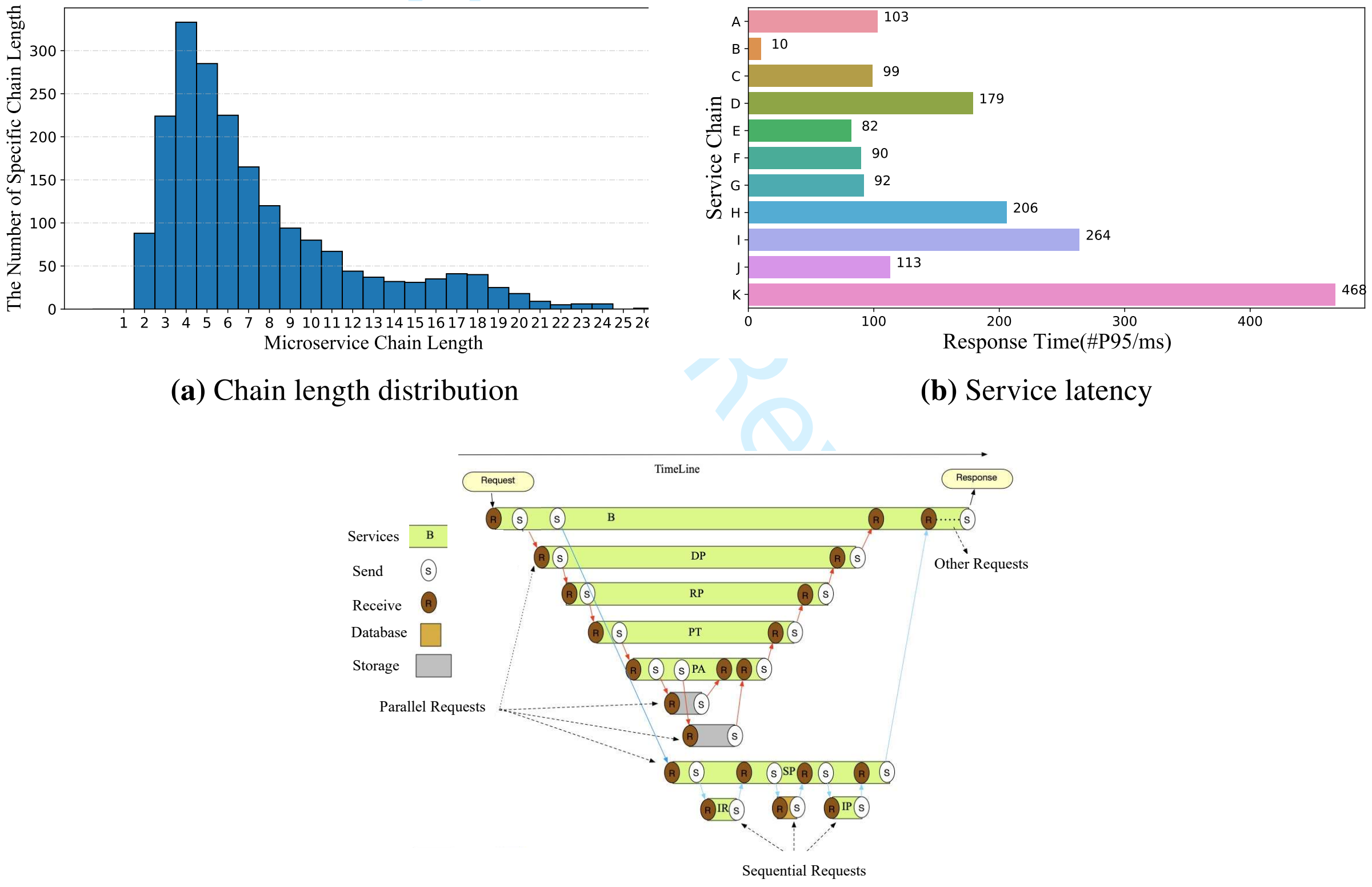

**(a)** Chain length distribution

**(b)** Service latency

**(c)** Execution History Graph

**Figure 2** Complex structure of microservice chains.

## 3.3 | Feature Analyses for Alibaba Microservices

From 2019, ASI has transformed all the core services into microservice-based applications and firmly served the largest Internet workloads in the world. To comprehensively understand the features of Alibaba's large-scale microservices and design optimized

[2]It is also called "traffic" in Alibaba scenario

resource provisioning policies, we first collect and analyze 15 TB raw data of Alibaba's microservice run-time information. We then summarize our key observations as follows:

**Observation 1 (*Ob*1): Chain length and service latency exhibit significant variability due to the intricate nature of microservice chains.** As depicted in Fig. 2a, the length of a typical commercial microservice chain can range anywhere from two nodes to over 20 nodes. Furthermore, as shown in Fig. 2b, the latency across various microservice chains can differ dramatically, with values ranging from 10 milliseconds to 400 milliseconds. This short latency necessitates the latency-critical microservice to perform faster than traditional cloud applications. Moreover, as demonstrated in Fig. 2c, the execution history graph depicts how requests can be executed across various microservices in either a sequential or parallel manner. These fluctuations render the behavior of microservice chains complex, making it challenging to provision adequate resources. For instance, excessive scaling of the resources can negatively impact the performance of the microservices.

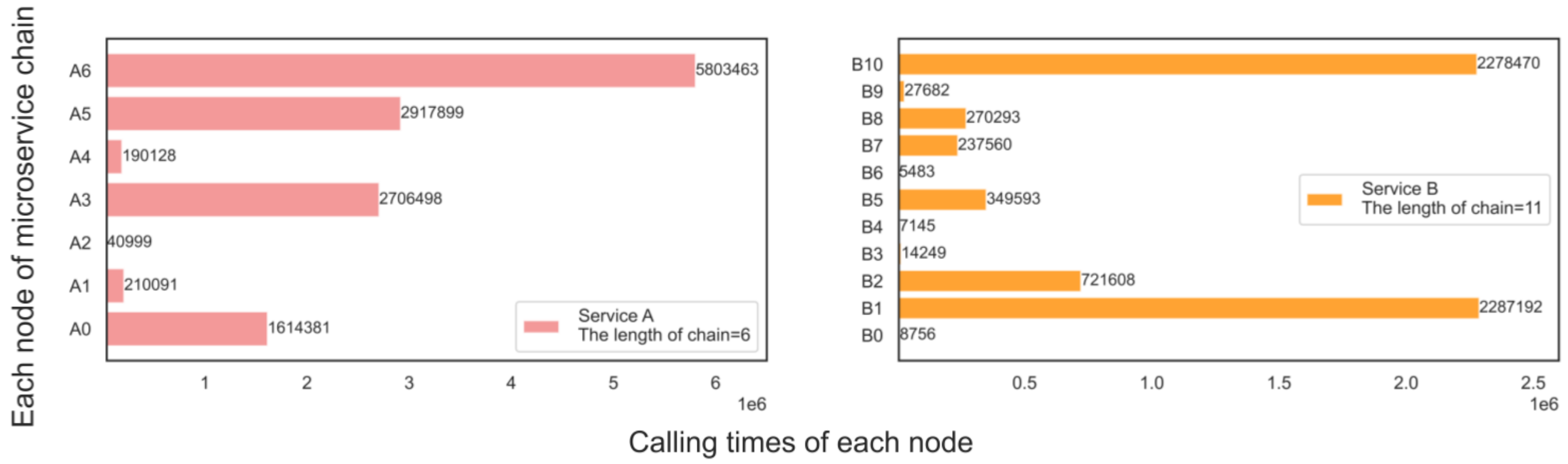

**Figure 3** Calling times of nodes based on different applications.

**Observation 2 (*Ob*2): Fluctuations in the frequency of calls lead to increased resource demands on specific nodes.** A microservice chain can comprise multiple microservices, and we have observed that the calling times of these microservices can vary significantly. As depicted in Fig. 3, we have selected two representative microservice chains in a business transaction scenario and found that even within the same microservice chain, the calling times of different microservices can vary by over five times. This observation highlights the importance of adequate resource allocation for the most frequently called microservice, such as node A6 in Service A or node B1 in Service B. If these nodes are not adequately resourced, it can result in a significant degradation of the performance of the entire microservice chain.

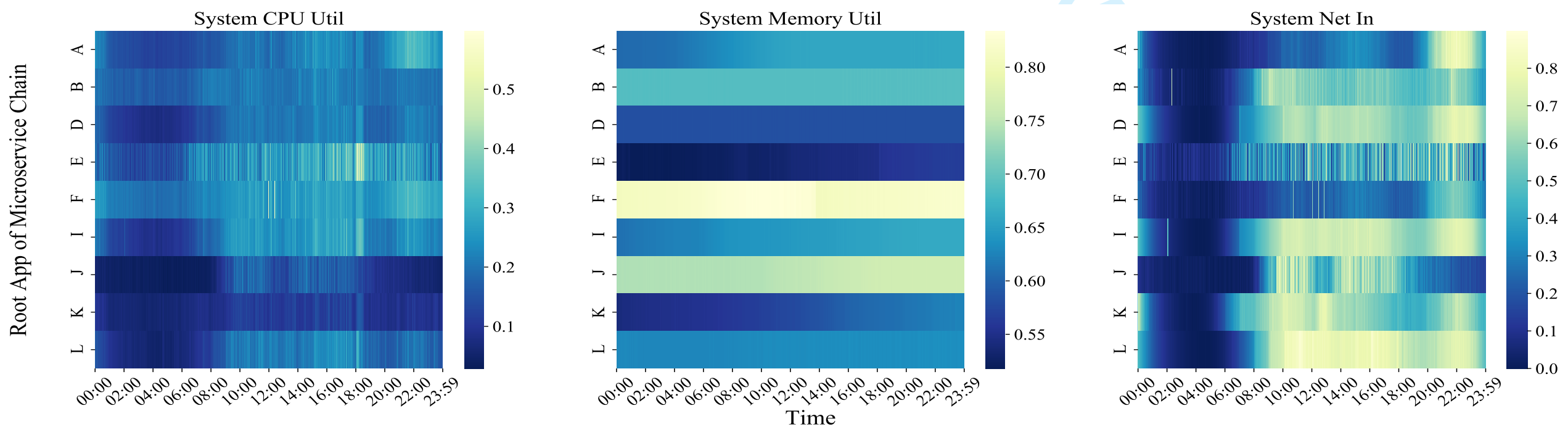

**Figure 4** Resource usage fluctuations via time in (a) CPU, (b) memory, and (c) network.

**Observation 3 (*Ob*3): The utilization of CPU and network resources fluctuates dynamically, while memory usage remains consistent.** The amount of resources required by each microservice instance is subject to change based on the volume of requests processed. To minimize costs, microservices are often co-located to promote efficient resource usage while avoiding resource contention. As illustrated in Fig. 4, the diagram demonstrates the resource usage during various time periods and shows that the utilization of CPU and network resources can vary among different microservices, while memory usage remains stable. CPU and network are the primary resource bottlenecks, while out-of-memory issues are rare in the Alibaba scenario.

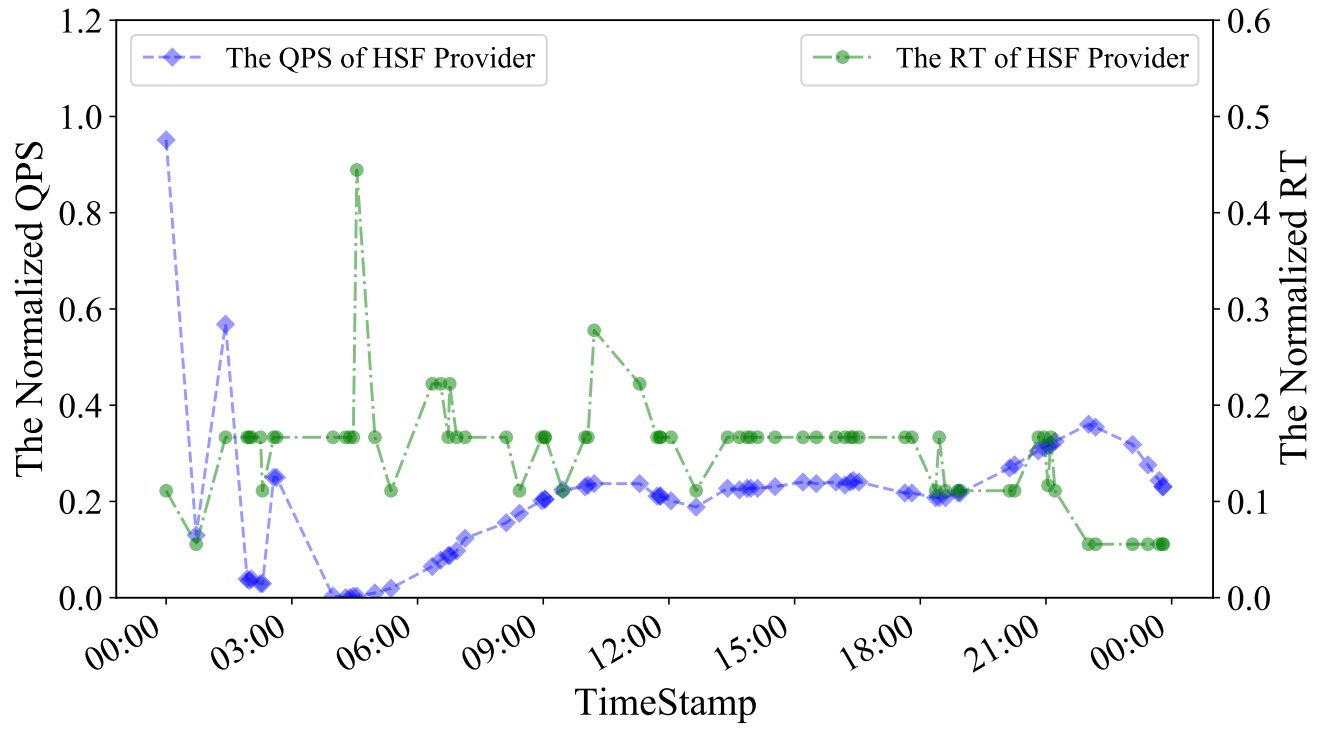

**Figure 5** QPS and response time fluctuations.

**Observation 4 (*Ob*4): An inconsistent relationship exists between QPS and response time.** Response time and QPS are crucial metrics for online business transactions, as depicted in Fig. 5, which illustrates the fluctuations in response time and variations in QPS. It is observed that the response time does not always follow the same trend as QPS. For example, it may decrease when QPS increases (e.g., at 6:00). This inconsistency can be attributed to resource mismatch. Based on an analysis of approximately 10,000 applications in Alibaba, we have observed that there is a misalignment of resources, such as CPU, memory, and Java virtual machine heap utilization. As shown in Fig. 6, 95% of the microservices consume less than 40% CPU utilization but higher memory utilization, 40% of the microservices require more than 80% memory, and 40% of the microservices can demand more than 60% heap utilization.

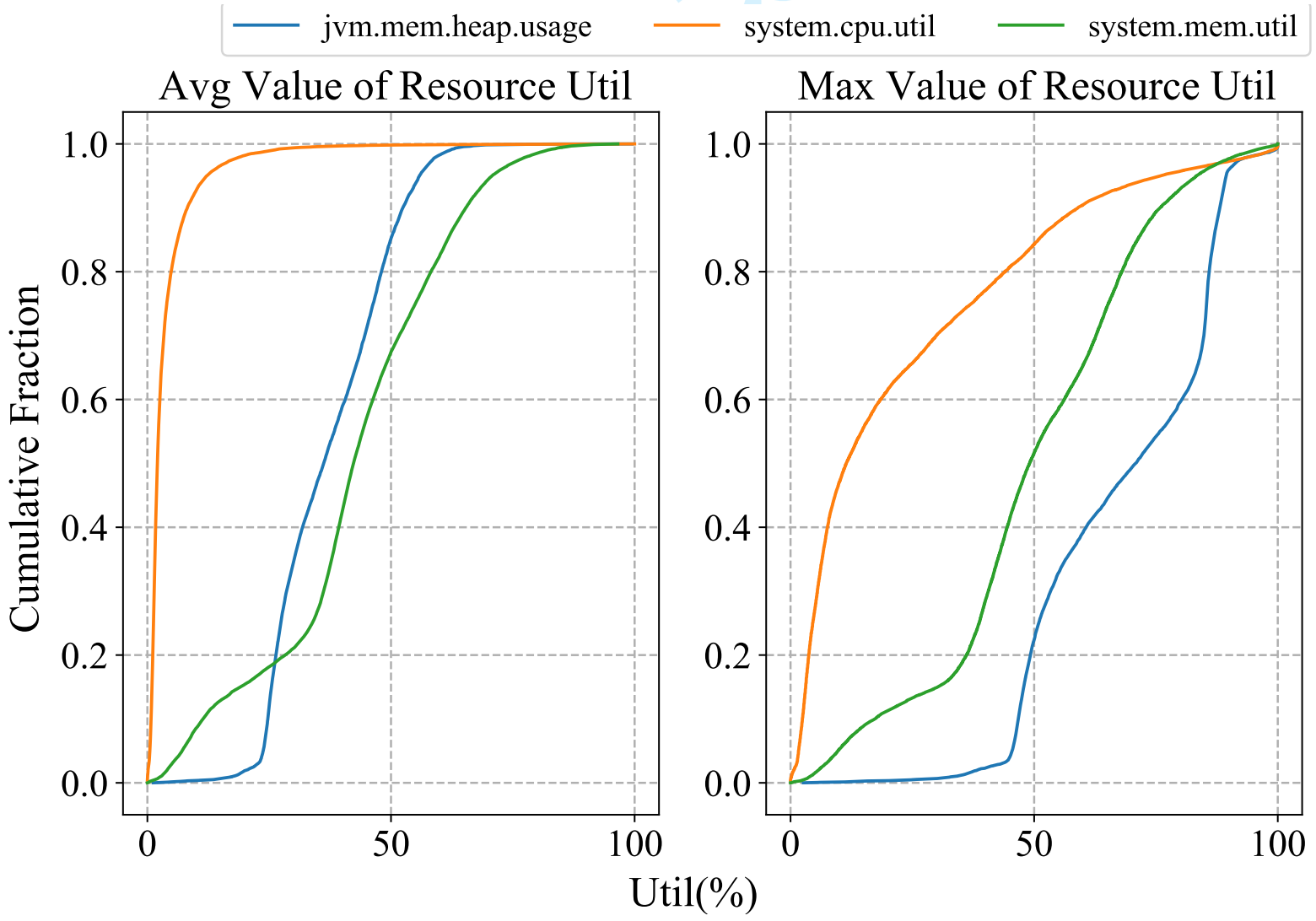

**Figure 6** Mismatched resource usage in CPU and memory.

## 3.4 | Challenges in Alibaba's Resource Provisioning

Based on the above observations, we can notice that the ASI faces heterogeneous nature of resources and tasks with different SLOs, the processing of massive workloads, and the management of microservices with complex properties, such as intricate service chains, fluctuating resource utilization, and mismatched resources. These have made several key **challenges** in efficient resource provisioning of Alibaba's microservices cluster as follows:

**Challenge 1 ($C1$): How to maintain low latency for latency-critical services in the face of an extra-large and heterogeneous hardware environment with numerous and complicated microservice applications?** Given the diverse types of microservice workloads, including online, offline, machine learning, and real-time stream processing, resource competition can lead to unpredictable fluctuations in QoS for online business, resulting in performance degradation, high tail latency, and even request failures. This is particularly true during high-load events such as the Double 11 Festival, when the large-scale cluster is more susceptible to increased latency. Despite the complexity of the environment, it is still crucial to maintain low latency for the microservice application, posing a challenge for resource capacity provisioning policies.

**Challenge 2 ($C2$): How to measure the performance interference under the co-location scenario given the extremely long and complicated chain of microservice applications?** When a performance issue occurs within a container, it can affect the performance of the service running within it, resulting in high tail latency[27]. To address this challenge, there is a need for clear, quantitative measurement of runtime QoS. Traditional monitoring methods may not be sufficient when handling millions of requests, and there is a need for a monitoring solution that is lightweight and can effectively handle dynamic scaling, node failures, load balancing, and failure migration. The solution should ensure transparent and accurate measurement of runtime QoS while reducing the burden on the system to maintain stability.

**Challenge 3 ($C3$): How to represent the capability of computation units and pinpoint resource bottlenecks?** To optimize the resource provisioning and system performance, the bottleneck of resource types should be identified. However, analyzing the relationship between the QPS and response time is challenging due to the vast number of configurable parameters and the complex relationship between the two metrics. This makes it difficult to clearly determine the cause of performance degradation. Therefore, accurately modeling the capacity of computational units to handle QPS is crucial, as it directly affects the ability to identify resource bottlenecks.

## 4 | RESOURCE PROVISIONING SOLUTION IN ALIBABA

In this section, we will present the basic resource management solution of Alibaba's current practice. The solution is designed based on following key principles:

**Unified Scheduler:** based on $Ob1$ and $C1$, the ASI needs to have the capability to provide scheduling for tasks with different priorities and QoS requirements with a unified scheduler. ASI provides scheduling and preemption mechanisms to over-provision resources, and cold memory can be recollected to enlarge the available resources of over-provisioning. More details will presented in Section 4.1.

**Fine-grained Management:** according to $Ob2$ and $C2$, co-located applications deployed on multiple containers can share the resources of different hardwares. ASI can restrict the CPU and memory resource usage of containers by blocking the calls of APIs to achieve fine-grained resource management. More details will be discussed in Section 4.2.

**History-based Resource Estimation:** as per $Ob3$, $Ob4$ and $C3$, the solution should predict the resource usage to optimize the resource provisioning. Therefore, the historical resource usage for statistical analyses should be collected, and the prediction and scheduling algorithms should be applied for workloads with different patterns. More details will be provided in Section 4.3.

### 4.1 | Workflow of Resource Capacity Platform

ASI offers a unified platform and workflow with comprehensive functions to manage and provision resources for Alibaba scenario, and its main workflow of the Capacity Platform is illustrated in Fig. 7. The *AHPA* (Advanced Horizontal Pod Autoscaler) is responsible for executing the decisions of Capacity Platform to provision resources based on microservice chain status, node utilization and various QoS requirements of applications to achieve the unified scheduling and optimize system performance. The *Metrics Collector* obtains the metrics of system and application status, and then stores them into the database (e.g. Mysql and HBase) for further use by other modules like *Data Processor*, *Application Filter* and *Algorithm Characterization*. The *Algorithm Filter* classifies the suitable applications based on resource usage pattern and workloads fluctuations to assign proper amount of resources. The characterization approach aims to mine the periodical tendency and generate the profile relationship between resource usage and performance indicators, which mainly contains rule-based characterization technique based on human maintenance experience, and benchmark characterization technique that records the number of containers used for the peak workloads during the last time period (e.g. Alibaba uses 35 days as baseline). Based on these models, the *Capacity Platform* can send the decision plan to the *AHPA* component according to a predefined trigger policy (e.g. threshold-based). The platform also provides the risk management and control mechanisms to block failures and networks, roll back to previous versions, and anomaly detection. In this work, we focus on discussing resource provisioning, therefore the detailed design of other components, like data recovery and data storage, are omitted.

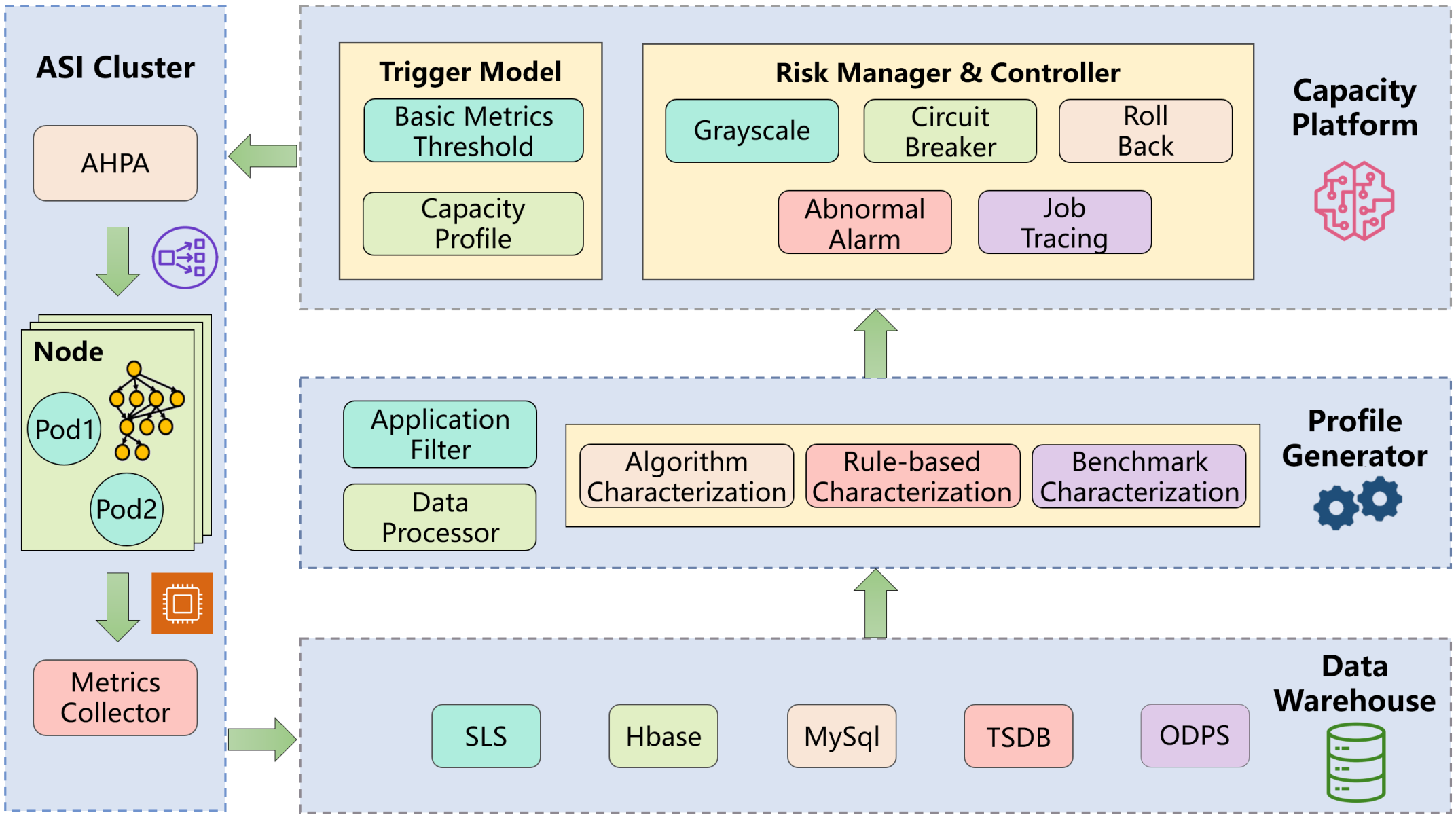

**Figure 7** Workflow of Resource Capacity Platform.

## 4.2 | Resource Provisioning Framework

Fig. 8 shows the general resource provisioning framework of ASI derived from the Kubernetes. Alibaba has integrated *AHPA* into the framework as an add-on to support the execution of resource provisioning actions. The *AHPA* accepts the resource provisioning decisions from the Capacity Platform as presented in Section 4.1, and interacts with the API Server that provides the unique entrance to conduct resource scheduling operations. *AHPA* manages the provisioned resources for pods in a fine-grained manager. It labels the pod replicates with traffic-on or traffic-off tags to decide whether admitting more workloads to specific pods or not. *AHPA* can also dynamically modify the number of pod replicates[3] via the *StatefulSet* component, which is aware of the resource updates of its associated pods, thus replicates can be created or deleted to reduce management risks. All fine-grained cluster resource usage data (e.g. pod level) is stored in the *ETCD* component that is a distributed key-value storage including configuration data, state data, and metadata. The *Walle* component is deployed on each node to collect the monitored metrics for nodes and pods, and the collected data is stored in the offline computation module to characterize resource capacity profile. Based on the collected data, the *Capacity Profile* module is responsible for searching available resources periodically, forecasting the future workloads, determining the capability of pod replicates to process QPS, and calculating the required resource capacity.

## 4.3 | Resource Provisioning Algorithms

The basic resource provisioning algorithms of Alibaba mainly consist of 5 key steps:

**Step 1: Data Collection.** The monitoring component plays a crucial role in collecting comprehensive metrics, such as QPS and response time, which are vital indicators of online business transaction performance. To ensure accurate and reliable data collection, ASI employs a sliding window approach with a window size of 5 minutes. The data is picked up at a second-rate to reduce the noise and minimize the risk of incorrect workload predictions. By configuring the maximum data of each sliding window as the data of the current window, the system can effectively mitigate the impact of fluctuating metrics and provide more stable and accurate results.

[3] In this work, the replicates refer to the pod replicates.

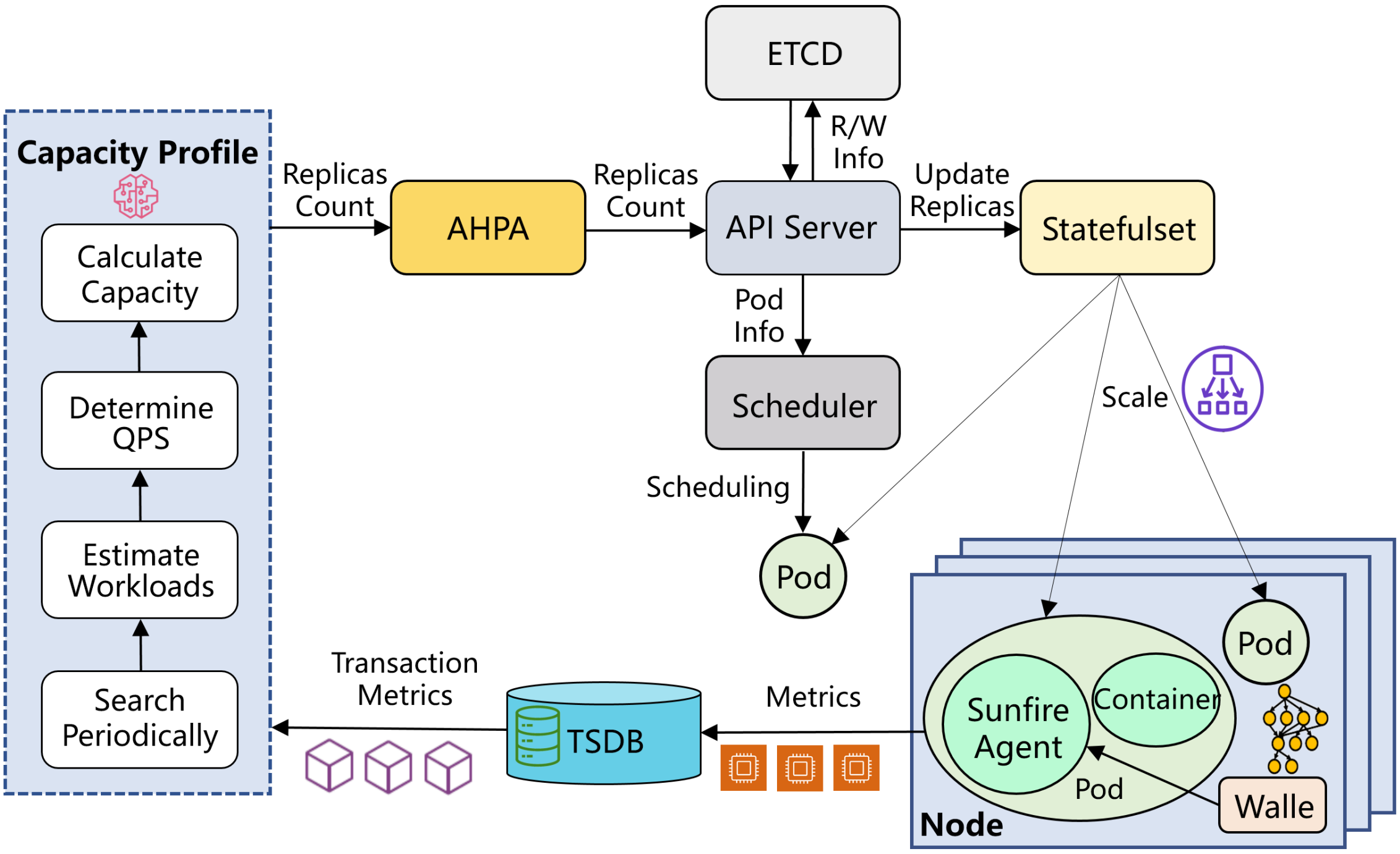

**Figure 8** Resource Provisioning Framework.

**Step2: Workloads Analyses and Estimation.** ASI also utilizes traditional machine learning techniques, such as LSTM and ARIMA, to predict the patterns of the workloads. This allows the system to effectively make use of spare resources, such as memory, as observed in the *Ob*3. To optimize the performance of these applications, ASI applies different resource provisioning policies depending on the type of workloads. For workloads that are periodical in nature, the system proactively makes scaling decisions based on the predicted workloads. On the other hand, for non-periodical workloads that cannot be predicted accurately, decisions are made reactively based on predefined thresholds.

**Step 3: QPS Modelling.** As highlighted in *Ob*4, establishing a direct correlation between QPS and response time is challenging. To address this, ASI utilizes a different approach to explore the relationship between QPS and provisioned resources. Instead of trying to fit the relationship between QPS and response time, ASI models the relationship between QPS and the computational capacity of resources based on pressure tests and practical experience. For example, using median QPS value as an indicator of a pod's capacity. This approach helps to maintain SLOs while optimizing resource utilization. The goal is to translate the workloads into the number of active pod replicas needed. With this information, ASI can determine and allocate the necessary number of pods to handle the workloads.

**Step 4: Proactive Resource Capacity Provisioning.** For workloads with periodical patterns (e.g. steady requests), proactive resource management policies can be implemented. ASI leverages a combination of machine learning techniques for workload prediction and empirical methods (such as provisioning the maximum number of historical replicas to avoid SLA violations) to determine the most appropriate scheduling actions for pods in the microservice calling chain.

**Step 5: Reactive Resource Capacity Provisioning.** For workloads that lack clear patterns and are difficult to predict (e.g. spikes), ASI employs a threshold-based approach inspired by the native Kubernetes HPA. This approach leverages predefined thresholds to make reactive scaling decisions for microservices.

# 5 | ENHANCED ALIBABA CAPACITY PROVISIONING (ALI-PRO) APPROACH

In this section, we propose our enhanced approach over Alibaba's current practice, named as Ali-Pro. The approach optimizes steps 2 to 5 of basic resource provisioning algorithms in Section 4.3.

## 5.1 | Workloads Analyses and Estimation

Under E-Business scenario, the workloads are closely relevant to the behavior of users. Therefore, most applications can have an apparent tendency during the observed period. It is reasonable to make workloads analyses to identify the typical features and estimate the future workloads. Based on the periodical feature, the workloads can be classified into two categories: periodical and non-periodical.

For applications with periodical pattern, we apply LightGBM [28], which is based on gradient boosting decision tree (GBDT) and has been validated to be efficient and accurate. The traditional GBDT approach used in Alibaba previously faces the challenges to balance the trade-offs between accuracy and efficiency as it needs to scan all the data for each feature to estimate the information gain of all possible points, which is not suitable for large-scale cluster due to high computation complexity. In addition, the scale of Alibaba's makes it unsuitable to use heavy deep learning approaches for prediction to ensure efficiency. LightGBM is light-weight and overcomes the limitations by combining gradient-based one-side sampling that randomly drops instances with small gradients and exclusive feature bundling that reduces effective features. To improve accuracy of model, we also apply grid search to the key parameters of decision trees including the maximum depth, number of leaf nodes, and learning rate. The statistical information such as mean, standard error, minimum value, maximum value, skewness, and kurtosis are obtained for analyses.

## 5.2 | QPS Modelling

After the workloads with apparent periodical trends are predicted by LightGBM, the workloads can be converted into resource usage based on the capability of instances to handle the variance of QPS. As at each moment, the Alibaba applications can be called by multiple middlewares and web requests sent from clients, the QPS is defined as the sum of QPS from web users and enterprise middlewares, where only the processed requests are counted. The workload data comes from long-term running data of online applications, and the configurations of provisioned resources are obtained from the maintenance experience.

We notice that the distribution of QPS tends toward a normal distribution, thus we apply regression-based approaches combining with the three-sigma rule, the QPS capability for a single instance can be modelled as $\mu + 3\sigma$ by removing anomaly data, where $\mu$ is the mean of the distribution and $\sigma$ is the standard deviation. In this way, the $\mu + 3\sigma$ value falls into the range of the dataset and has been evaluated in a realistic environment. With the load balancer in the system to distribute the requests evenly to different instances, this approach can efficiently avoid SLA violations in practice.

## 5.3 | Proactive Resource Capacity Provisioning

Based on our workloads prediction and QPS modeling techniques, proactive resource capacity provisioning for managing the number of pod replicates can be conducted to handle workloads with periodical patterns. These actions can inform the scaling module about how many replicates should be provisioned during different periods. By extending the empirical practice of Alibaba, we propose Algorithm 1 as finding the shortest path in the directed acyclic graph, where the shortest path represents the minimum number of required replicates.

Algorithm 1 shows the shortest path based approach to achieve the objective that ensures system performance with the minimum resources. The algorithm firstly converts the number of replicates at different time slot (e.g. each slot is one hour and 24 hours in total) into nodes in a graph. Then it initializes all nodes with infinite distance and the source node is initialized with 0 (lines 1-3). The distance between different nodes is initialized by the LightGBM prediction algorithm, for instance, distance with value of 5 means that 5 more replicates should be added. After that, the algorithm calculates the distance of all neighbor nodes and finds the node with the smallest distance. The selected node with the minimum distance is put into the solution set as an element. The operations continue until all the nodes have been checked (lines 7-12), where the final solution $S$ contains the minimum number of replicates should be provisioned at different time slots. The solution can be implemented based on priority queue (the head of the queue is with the minimum value) and the algorithm complexity is $O(E + VlogV)$, where $E$ is the number of edges in the graph and $V$ is the number of the vertex.

**Algorithm 1:** Proactive Capacity Provisioning

**Input:** Directed acyclic graph $G$ with node $v \in G$, $pre[v]$ stores the previous node of $v$, $dist[u]$ stores predicted number of provisioned replicates of node $u$, $dist[v, u]$ stores the changed number of replicates by prediction, solution set $S$, neighbour nodes $N(u)$ of node $u$.

**Output:** Solution $S$ containing provisioned number of pod replicates

**for** $v \in G$ **do**
  $dist[v] = \infty$
  $pre[v] = \emptyset$
$S = \{s\}$
$dist[s] = 0$
**while** $G \neq \emptyset$ **do**
  $u$ = node in $G$ with the smallest dist[]
  $G = G \setminus \{u\}$
  **for** $v \in N(u)$ **do**
    **if** $dist[u] + dist[u, v] < dist[v]$ **then**
      $dist[v] = dist[u] + dist[u, v]$
      $S = S \bigcup u$

## 5.4 | Reactive Resource Capacity Provisioning

By extending Alibaba's threshold-based reactive resource capacity provisioning, we propose an algorithm based on resource utilization threshold to adjust the number of pods. Algorithm 2 shows the pseudocode of reactive resource capacity provisioning with periodical analyses. There are two principles in designing the algorithm: a) configuring threshold with safety parameter to avoid the request bursts; b) configuring cooling time to avoid high costs incurred by frequent scaling operations.

The algorithm firstly initializes the system parameters, including the size of the sliding window, the predefined maximum and minimum number of replicates, and other configurable parameters. In the scheduling time interval $T$, the algorithm calculates the statistical data of metrics in each time interval $t$ (lines 1-2). The statistical data include the mean value, the maximum value, or the 95 percent value of metrics like response time and CPU utilization. Then the algorithm computes the scaling parameter based on the predefined scaling threshold $t^*$ and the current metric data (line 3). If the scaling parameter is within the safety area, the predicted number of replicates $\hat{C_{t+\sigma}}$ can be obtained through the calculated parameter from previous steps and the number of replicates at the current time interval (line 4). Compared with the maximum and the minimum number of replicates, the predicted number of replicates will be updated (lines 5-12). To avoid too frequent scaling with high costs, a cooling time of removing replicates is also defined. If the previous time is within the cooling time, then the replicates removal operations will not be executed (lines 13-16). Finally, the expected number of replicates can be decided as $C_{t+\sigma}$.

# 6 | PERFORMANCE EVALUATIONS

In this section, we provide a detailed overview of the experimental setup and the metrics we utilized for evaluations. Subsequently, we present the results obtained from Alibaba's cluster using our proposed Ali-Pro approach compared with several baseline methods.

## 6.1 | Experiment Setup

In this subsection, we will introduce the cluster used for our experiments, along with the state-of-the-art baseline approaches we have chosen for comparison. Additionally, we will outline the metrics we employed to evaluate the performance of these approaches.

**Algorithm 2:** Reactive Capacity Provisioning

**Input:** Sliding window size $W$ to measure the statistical data, time interval $t$, the maximum number of replicates $C_{max}$, the minimum number of replicates $C_{min}$, the number of replicates $C_t$ at time interval $t$, the safety parameter for scaling $sp \in [0, 1]$, cooling time $ct$ of removing replicates, threshold $t^*$, statistical data $SD$ for specific metric, e.g. response time $rt$, a period of time interval $\sigma$.

**Output:** Number of replicates $C_{t+\sigma}$

**for** $t = 1$ *to* $T$ **do**
 $rt(t) \leftarrow SD_{rt}(rt(\tau)|\tau \in [t - W, t])$
 $n_r(t) \leftarrow rt(t)/t^*$
 $\hat{C_{t+\sigma}} \leftarrow n_r(t) * C_t$
 **if** $|n_r(t) - 1| > sp$ **then**
  **if** $C_{min} < \hat{C_{t+\sigma}} < C_{max}$ **then**
   $C^*_{t+\sigma} = \hat{C_{t+\sigma}}$
  **else**
   **if** $C_{min} \geq \hat{C_{t+\sigma}}$ **then**
    $C^*_{t+\sigma} = C_{min}$
   **else**
    $C^*_{t+\sigma} = C_{max}$
 **if** $C^*_{t+\sigma} < C_t$ & $\delta < ct$ **then**
  $C_{t+\sigma} = C_t$
 **else**
  $C_{t+\sigma} = C^*_{t+\sigma}$

### 6.1.1 | Cluster and Baselines

To evaluate the performance of improving resource capacity provisioning, we have conducted our experiments on cluster of ASI. The physical nodes are with Intel Xeon Platinum 8163 CPU @2.50 GHz and Platinum 8269CY CPU @ 2.5 GHz. The workloads are derived from a typical E-Business application that we name as application A whose features are demonstrated in Section 3.3. The application is distributed on homogeneous virtual machines with 104 cores, 512 GB memory, and 320 GB disk. The instances of application A are running in pods, each pod can require resources with a maximum 4 cores CPU, maximum 8 GB memory, and maximum 64 GB disk.

Several baselines have been used for comparison with our extended approach **Ali-Pro**:

**Over-Pro** is the static algorithm that always keeps the maximum number of replicates by resource over-provisioning, which can assure QoS.

**Kube-Pro**[29] is derived from the society version of Kubernetes. It has been optimized and used in Alibaba with the mechanisms introduced in Section 4.3.

**Optimal-Pro**[16] is a theoretical number of replicates based on the amount of workloads coming into the system, which is a white box approach that can reach optimal results.

**Conserv-Pro** is a conservative approach based on Ali-Pro to allocate more resources, where the time to make scaling operations has been shifted to a longer period than Ali-Pro, i.e., 30 minutes earlier to add replicates and 30 minutes later to remove replicates than Ali-Pro.

### 6.1.2 | Metrics

Apart from the widely used metrics like resource utilization and latency[30] to validate elasticity, the following metrics have been adopted to evaluate prediction accuracy and resource usage:

**1) RMSE:** it represents the root mean square error that can measure the difference between the predicted and actual workloads, which is defined as:

$$RMSE = \sqrt{\frac{1}{n}\sum_{i=1}^{n}(y_i - \hat{y}_i)^2} \quad (1)$$

where $n$ is the size of data, $y_i$ is the actual data, and $\hat{y}_i$ is the predicted data.

**2) MAPE:** it is another widely used metric to measure prediction errors as:

$$MAPE = \frac{1}{n}\sum_{i=1}^{n}|\frac{y_i - \hat{y}_i}{y_i}| \quad (2)$$

**3) SP:** one major concern of Alibaba is to improve resource usage, and this metric is used to evaluate the percentage of saved resources, $T$ is the period that indexed by $i$ and $j$ to scale resources, and $R$ is the number of replicates in each period:

$$SP = 1 - \frac{\sum_{i=1} R_i * T_i}{\sum_{j=1} R_j * T_j} \quad (3)$$

## 6.2 | Workloads Estimation Results

In order to evaluate the accuracy of the workloads prediction, we compare LightGBM with several ML-based approaches for time-series data prediction, including LSTM, ChatBoost, Random Forest and ARIMA[31]. Four applications derived from online shopping are evaluated. Table 1 shows the RMSE and MAPE for the applications, which shows the LightGBM based approach can obtain the best prediction results in applications A and B. Although for Applications C and D, the prediction results are not the best, the LightGBM still outperforms LSTM and Random Forest, and achieves close results with the best approach, like ARIMA. The main reason is that LightGBM can produce much more complicated trees by following leaf wise split approach to achieve higher prediction accuracy.

**Table 1** The evaluations of workloads prediction models

| | LSTM | LightGBM | CatBoost | Random Forest | ARIMA |
|---|---|---|---|---|---|
| | Application A | | | | |
| RMSE | 5603.78 | **1255.57** | 1460.36 | 6949.45 | 12966.83 |
| MAPE | 0.25 | **0.10** | 0.11 | 0.59 | 0.63 |
| | Application B | | | | |
| RMSE | 516.89 | **2671.97** | 3323.82 | 6949.45 | 5231.03 |
| MAPE | 0.12 | **0.11** | 0.12 | 0.15 | 0.21 |
| | Application C | | | | |
| RMSE | 472.65 | 191.335 | **143.41** | 639.69 | 491.47 |
| MAPE | 0.27 | 0.13 | **0.12** | 0.59 | 0.45 |
| | Application D | | | | |
| RMSE | 982.93 | 697.61 | 742.76 | 1814.86 | **650.63** |
| MAPE | 0.163 | 0.13 | 0.12 | 0.274 | **0.108** |

## 6.3 | QPS Modelling for Pods

To investigate the pattern of QoS modelling for pods, We use the statistical data of application A for analyses, and other applications can be analyzed in the same way. To obtain the threshold of the single instance with corresponding QPS, we select the data generated between 8:00 to 20:00, as the workloads outside of this period is much lower than the observed period. We collect 442,686 data about the processed QPS by each instance within 11 days, and the observed distribution is demonstrated in Fig. 9. The red line is the fitted normal distribution based on the collected data, where the mean value $\sigma = 573.7$ and standard deviation value $\mu = 65.9$. Therefore, the threshold of QPS for a pod can be configured as $\mu + 3\sigma = 771.4$ based on the three-sigma rule. With this modelled value and total number of requests, the required number of replicates can be computed to optimize resource usage (e.g. obtaining the demanded number of replicates).

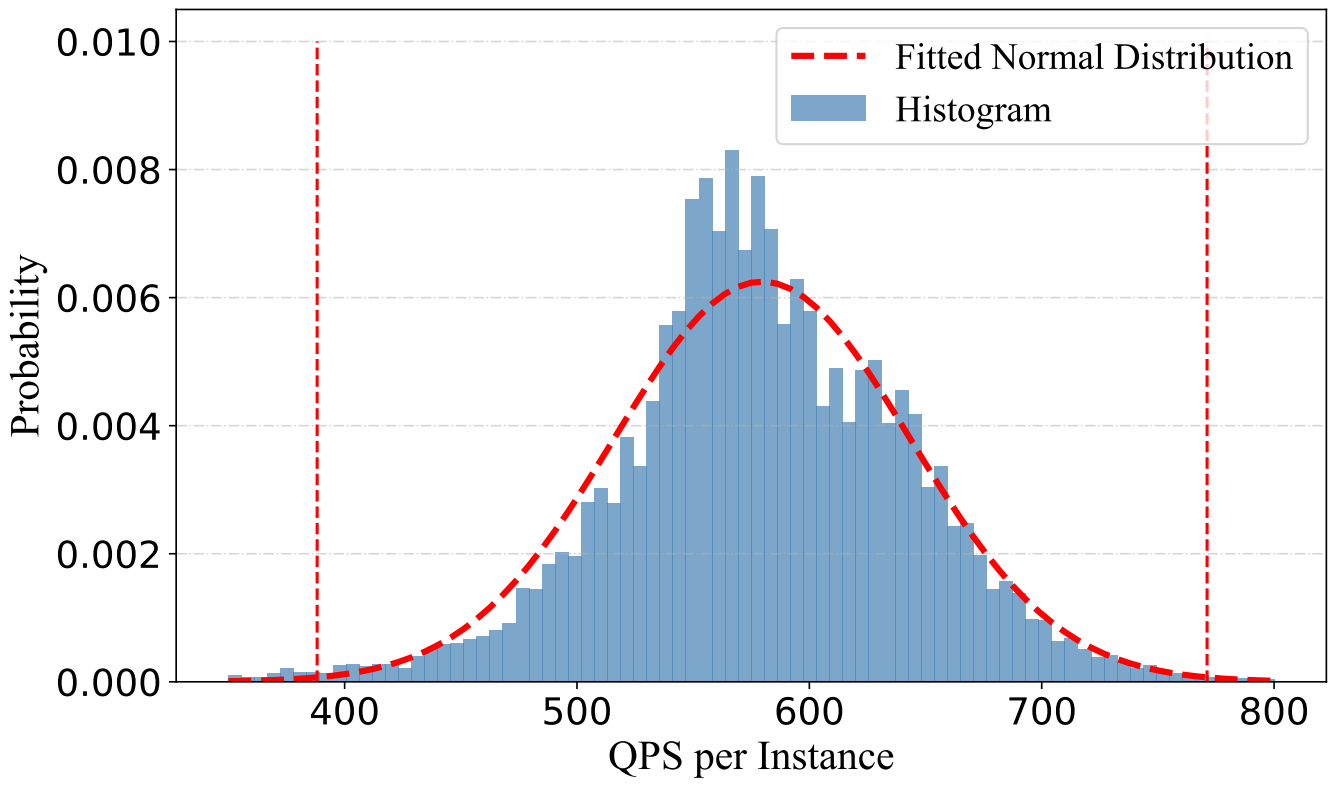

**Figure 9** QPS distribution of pods.

## 6.4 | Resource Capacity Provisioning Performance Analyses

Fig. 10 illustrates the number of used replicates based on different approaches, which can represent the resource utilization. We can notice that our proposed Ali-Pro can achieve better performance than other baselines except the Optimal-Pro with the theoretical results, which is hard to realize in practice. Over-Pro keeps the maximum number of replicates as 18 by over-provisioning, which leads to the maximum amount of wastage resources. Conserv-Pro based on a sliding window can have more pods than required during midnight, as its decision is based on results in 15:00 to 24:00 that with a large number of pods. Kube-Pro does not perform well as it is mainly based on the static threshold. With Over-Pro as the baseline to calculate $SP$ metric in Equation (3), considering Over-Pro as 0.0, then Conserv-Pro is 15.9%, Kube-Pro is 13.3%, Optimal-Pro is 24.0% and Ali-Pro is 18.1%. Moreover, the proposed approach performs well during the period from 0:00 to 7:00 with low workloads, where the $SP$ of Conserv-Pro is around 80.0%.

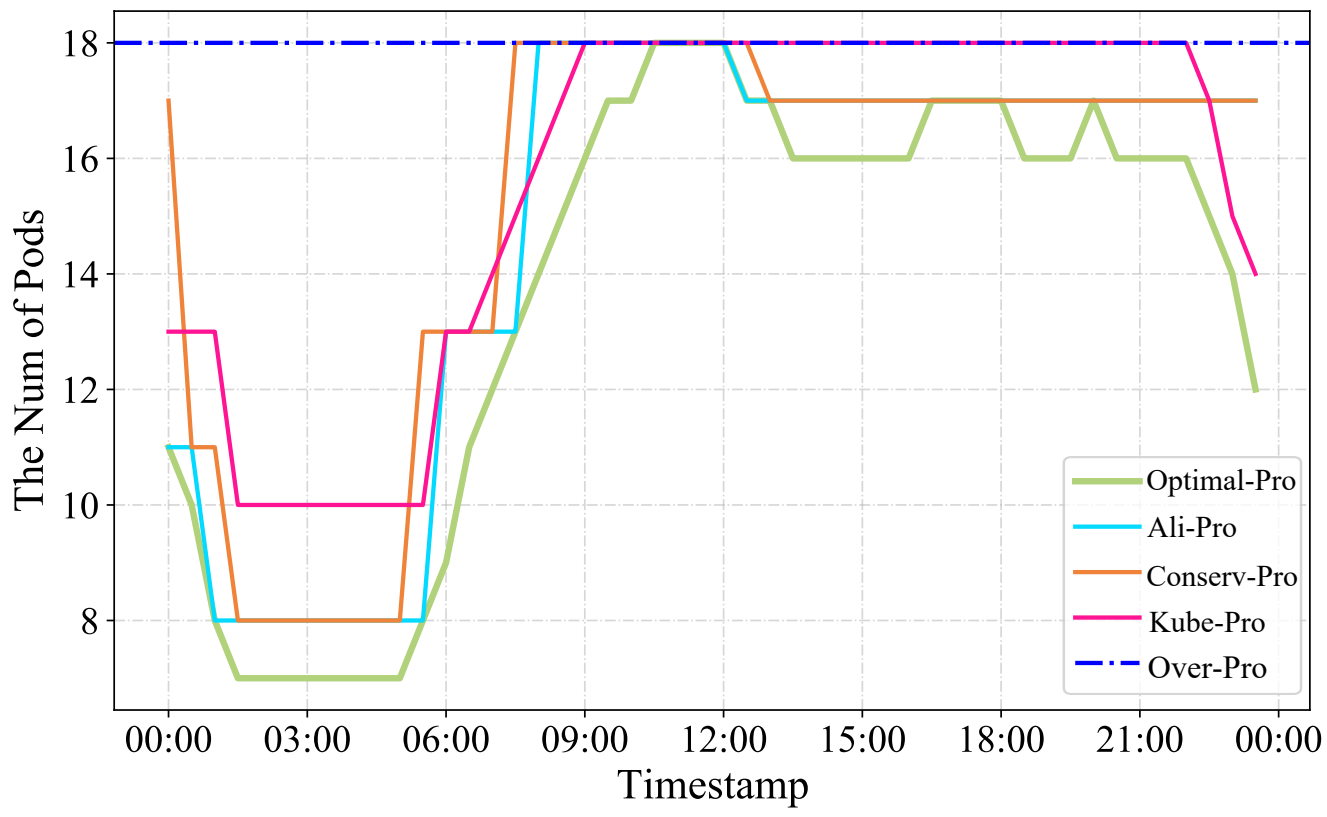

**Figure 10** Number of used pod replicates.

As Ali-Pro and Conserv-Pro have demonstrated good performance in saving replicates, we also investigate their performance on assuring QPS. Since Conserv-Pro and Ali-Pro can achieve close results, to avoid the overlapping of results on the same figure, we only use Conserv-Pro for comparison. Additionally, Conserv-Pro is more conservative than Ali-Pro, and the industry is more prone to use conservation approach to provide stable user experience. Fig. 11a demonstrates that Conserv-Pro can achieve the close performance with Over-Pro, and Fig. 11b shows Conserv-Pro can improve the CPU utilization of pods from about 10% to 15% compared with Over-Pro. The utilization is not high as the left resources on the pods can be used for other types of services, e.g. batch and best efforts.

We also conduct experiments to compare the application latency of Conserv-Pro and Over-Pro. Over-Pro can assure QoS as resources are over-provisioned, therefore, we compare Conserv-Pro with Over-Pro. Fig. 12a shows the latency distribution

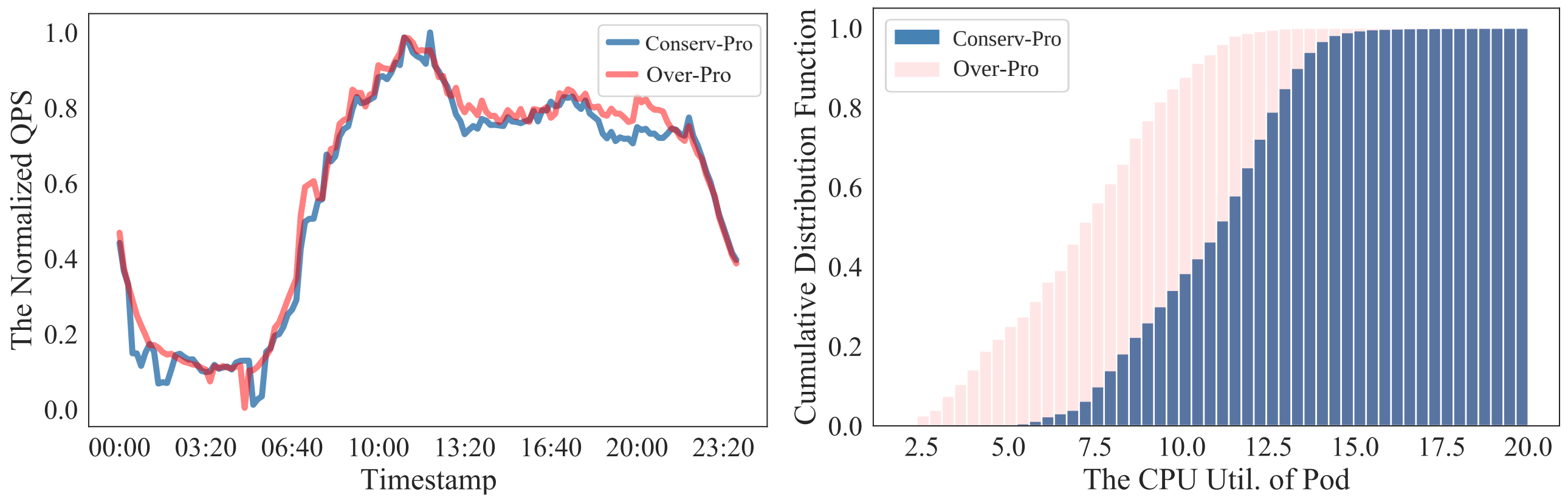

**(a)** QPS results **(b)** CPU utilization of pods comparison, and the results of Conserv-Pro demonstrate it can achieve close latency performance with Over-Pro, which means the performance on latency can also be ensured by Conserv-Pro. Fig. 12b represents the load distribution, which shows that the Conserv-Pro can also keep a similar distribution with Over-Pro. We also use Kullback-Leibler divergence (KL)[32] to measure how the distribution is different from the other, the lower KL value represents a closer relationship. Based on the distributions, the KL values of latency and load are 0.025 and 0.227 respectively, which validate the closeness of the distributions.

**Figure 11** QPS and utilization comparison.

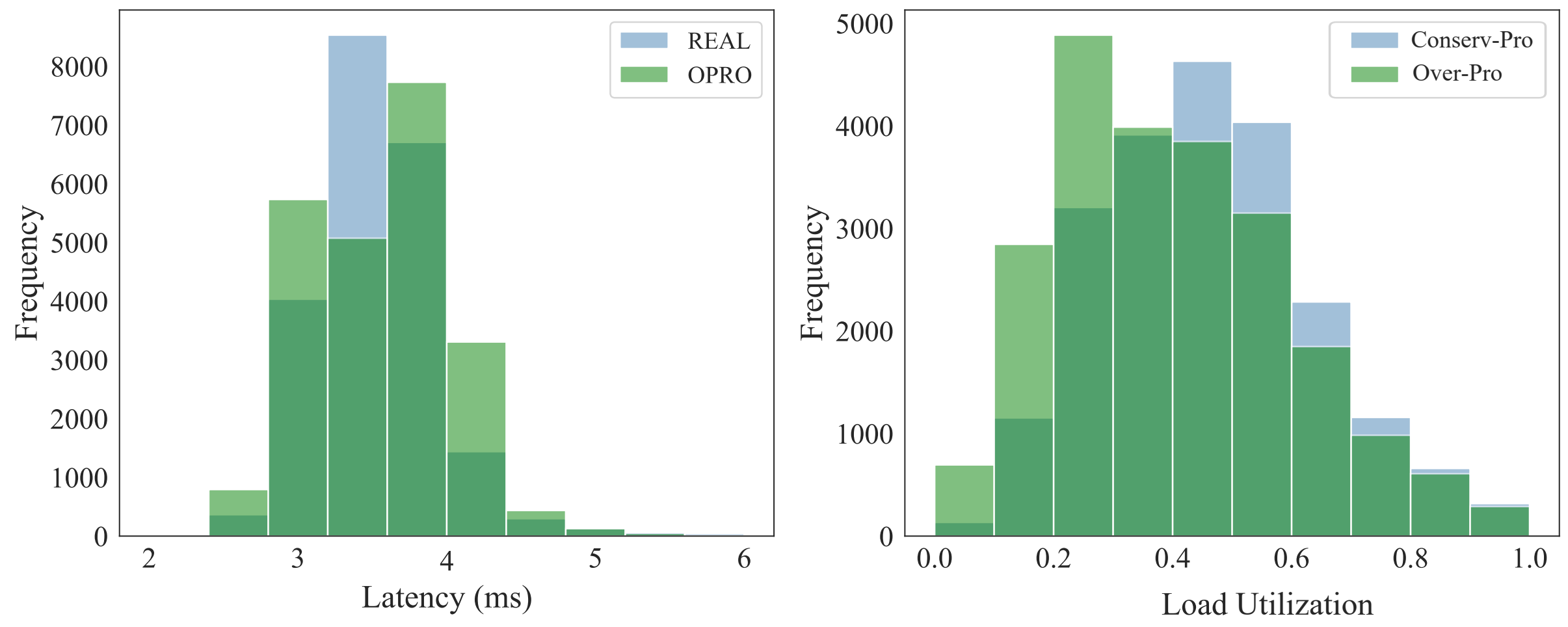

**(a)** Latency distribution **(b)** Load distribution

**Figure 12** Response time and load distribution.

In conclusion, our enhanced approach can optimize the number of used replicates over Alibaba's current practice while ensuring QPS and latency. This can benefit Alibaba to improve its revenues while assuring user experience.

## 7 | CONCLUSIONS AND FUTURE WORK

This work introduces Alibaba's extensive expertise in providing resource capacity for large-scale microservices clusters. The ASI platform, equipped with a unified scheduler, fine-grained scheduling, and history-based resource estimation, has effectively optimized resource utilization in microservice systems and has been instrumental in supporting Alibaba's core business for several years. To enhance our understanding of microservices, we conduct thorough analyses of microservices clusters, emphasizing key observations and challenges related to resource provisioning.

Based on these observations, we employ ML-based workload prediction, QPS modeling, as well as proactive and reactive resource provisioning techniques to tackle various workload patterns and address the identified challenges. We introduce resource provisioning algorithms capable of improving resource utilization by 10-15% while maintaining QPS and latency for microservice-based clusters. This work contributes to both academic and industrial knowledge by enhancing our understanding of large-scale microservice clusters, designing general resource provisioning frameworks for production environments, and optimizing resource provisioning for microservices.

For future research, we aim to investigate complex aspects of microservice dependency in large-scale clusters to further enhance resource usage efficiency. Additionally, we plan to explore the interference between different microservices in co-location environments, particularly for deep learning applications where various deep learning or machine learning tasks are deployed on heterogeneous CPUs and GPUs, necessitating the assurance of performance.

## ACKNOWLEDGMENT

This work is supported by National Key R&D Program of China (No.2021YFB3300200), the National Natural Science Foundation of China (No. 62072451, 62102408), Shenzhen Industrial Application Projects of undertaking the National key R & D Program of China (No. CJGJZD20210408091600002), Shenzhen Science and Technology Program (No. RCBS20210609104609044) and Alibaba Innovative Research Program.

□